\documentclass[preprint,12pt]{elsarticle} 
\usepackage[T1]{fontenc}
\usepackage[utf8]{inputenc}
\usepackage{graphicx}
\usepackage{amssymb}
\usepackage{hyperref}
\usepackage{longtable}
\usepackage{pdflscape}
\usepackage{multirow, booktabs}
\usepackage{multicol}
\usepackage{graphicx,type1cm,eso-pic,color}
\usepackage{tikz} 
\usepackage{graphicx}
\usetikzlibrary{positioning}
\usepackage{fancyhdr}
\usepackage[all]{hypcap}
\usepackage{lmodern}
\usepackage{textcomp}
\usepackage{multicol}
\usepackage{subcaption}
\usepackage{dcolumn}
\usepackage{epsfig}
\usepackage{amsmath,amsthm}
\usepackage{amsfonts,amssymb}
\usepackage{float}

\newtheorem{de}{Definition}[section]
\newtheorem{theo}{Theorem}[section]    

\newtheorem{prop}{Proposition}[section]

\journal{Epidemics}

\begin{document}

\begin{frontmatter}


\title{Mathematical Model for Transmission Dynamics of Tuberculosis in Burundi}



\author[1]{Steve Sibomana}
\author[1,2]{Kelly Joelle Gatore Sinigirira\corref{mycorrespondingauthor}}
\cortext[mycorrespondingauthor]{Corresponding author}
\ead{kelly.gatore@ub.edu.bi}
\author[3,4]{Paterne Gahungu}
\author[1,2,5]{David Niyukuri}

\address[1]{Department of Mathematics, Faculty of Science, University of Burundi, Bujumbura, Burundi}
\address[2]{Doctoral School, University of Burundi, Bujumbura, Burundi}
\address[3]{Institute of Applied Statistics, University of Burundi, Bubanza, Burundi}
\address[4]{African Institute for Mathematical Sciences (AIMS), Rwanda}
\address[5]{The South African Department of Science and Technology--National Research Foundation (DST-NRF) Centre of Excellence in Epidemiological Modelling and Analysis (SACEMA), Stellenbosch University, Cape Town, South Africa}

\begin{abstract}
Tuberculosis (TB) is among the main public health challenges in Burundi. The literature lacks mathematical models for key parameter estimates of TB transmission dynamics in Burundi. In this paper, the supectible-exposed-infected-recovered (SEIR) model is used to investigate the transmission dynamics of tuberculosis in Burundi. Using the next generation method, we calculated the basic reproduction number, $R_{0}$. The model is demonstrated to have a disease-free equilibrium (DEF) that is locally and globally asymptotically stable. When the corresponding reproduction threshold quantity approaches unity, the model enters an endemic equilibrium (EE). That means, the disease can be controlled through different interventions in Burundi. A sensitivity analysis of the model parameters was also investigated. It shows that the progression rate from latent to becoming infectious had the highest positive sensitivity, which means that $R_{0}$ increases and decreases proportionally with an increase and a decrease of that progression rate. 
\end{abstract}

\begin{keyword}
Tuberculosis, Reproduction number, disease free-equilibrium, endemic equilibrium, Lyapunov function, Burundi
\end{keyword}

\end{frontmatter}

\section{Introduction}

Tuberculosis (TB) is an airborne disease caused by Mycobacterium tuberculosis (MTB)\cite{T2}, and it is reported to be the second leading cause of morbidity and mortality in the world from a single infectious agent after the human immunodeficiency virus (HIV)\cite{T1}. The MTB  spreads through inhaling droplets from the cough or sneeze of a person suffering from active tuberculosis. The bacteria enters the body causing an MTB infection affecting mainly the lungs, but it can also affect any other part of the body including the urinary tract, brain, lymph nodes, bones, joints and the ear. Person with lowered immunity such as those with HIV, diabetes, immune disorders, and stage renal disease, those on drugs that suppress immunity, young children, pregnant women among others are at a higher risk of contracting the disease\cite{T3,T4}.

In Burundi, TB remains a public health problem, it is rampant in an endemo-epidemic mode and all layers of the population are concerned. In general, infection with tuberculosis is very likely to be asymptomatic for healthy people. The lifetime risk of developing clinically active TB after being infected is about $10\%$ \cite{T6}. People who have latent TB infection are not clinically ill or capable of transmitting TB \cite{T6,T7}. The immunity of older people who have previously been infected may decrease, and they may then be at risk of developing active TB of either exogenous reinfection (that means acquisition of a new infection from another infectious individual) or an endogenous reactivation of latent bacilli (that means the reactivation of a pre-existing dormant bacillus infection)\cite{T8}. Latent and active tuberculosis can be treated with antibiotics. However, its treatment has a side effects (sometimes quite serious) and takes a long time.

Carriers of the tuberculosis bacillus who have not developed tuberculosis disease can be treated with only one Isoniazid, also known as isonicotinic acid hydrazide (INH). For active tuberculosis it is often used together with rifampicin, pyrazinamide, and either streptomycin or ethambutol\cite{T9}. Unfortunately, it should be taken religiously for 6-9 months. Treatment of people with active tuberculosis requires simultaneous use of three drugs for a period of at least 12 months. Lack of compliance 	with these drug treatments, do not only can cause a relapse, but the development of antibiotic resistant tuberculosis, one of the most serious public health problems facing today's society. 

Studying the spread of TB using statistics and mathematics models did not receive enough attention in Burundi. As a result, we only observed a very limited use of mathematical models in studying the dynamics of TB transmission in Burundi's human populations. In the broad scientific literature, communicable diseases such as measles, influenza, rubella, among others, have been studied by many mathematical models \cite{T10,T11,T12}. These diseases have a number of traits, such as the fact that they frequently create epidemics and that transmission rates are greatly influenced by age-dependent contact rates. The etiological agents of these communicable diseases are viruses of different families but all capable of generating similar symptoms. Waaler and Anderson were the first who modeled mathematically tuberculosis transmission dynamics\cite{T13}.

In this work, a mathematical model for the transmission dynamics of TB in Burundi has been developed. We determine the existence and positivity of the system, and we provide the disease's equilibrium points and the reproduction number. The stability of the disease equilibrium is then determined. A sensitivity analysis is performed on the model parameters. Based on the TB data from Burundi, simulations and interpretations of the results will be carried out.

\section{Methods}
\subsection{Mathematical model}

\begin{figure}[!h]
	\centering
	\includegraphics[scale=0.55]{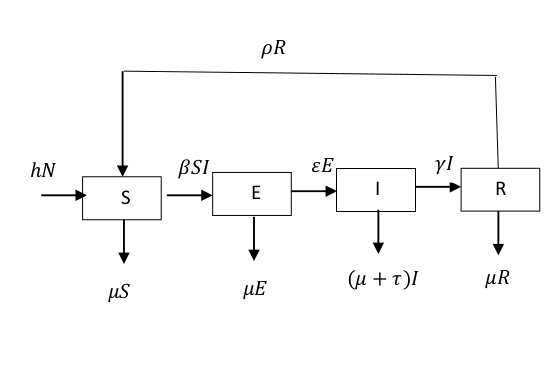}
	\caption{Diagram of the Susceptible-Exposed-Infected-Recovered (SEIR) Model}
 \label{f1}
\end{figure}

A mathematical model is be established using Susceptible-Exposed-Infected-Recovered (SEIR) compartmental approach as shown on  Figure \ref{f1}, where $S(t)$, susceptible humans, $E(t)$, exposed humans to tuberculosis, $I(t)$, infected humans with active tuberculosis,  $R(t)$, recovered humans at time $t \geq 0$. We consider that susceptible population keep increasing by the human birth at  $h$ rate. Thee loose of immunity is considered to be at rate $\rho$, and there is no permanent immunity to tuberculosis. Humans can contract MTB tuberculosis through contact with individuals who are infected with the disease. Therefore, they enter the exposed (latent) with $\beta$ rate. A proportion of the exposed class develop active tuberculosis, thus, moving into the infectious class with $\varepsilon$ rate of progression. If treatment is administered promptly, those who recover from the disease will move to the recovered class at $\gamma$ rate. Each human class decreases also by natural mortality at $\mu$ rate, except the infectious class which adds the mortality caused by the MTB tuberculosis at $\tau$ rate. $N(t)$ is used to denote the total population at time $t$ such that $N(t)=S(t)+E(t)+I(t)+R(t)$. All parameters of the model are assumed to be non-negative and the total human is assumed to be constant.


From Figure \ref{f1}, we can write the following system of ordinary differential equations:

\begin{equation}\label{e1}
\left\lbrace 
	\begin{array}{ccc}
	\frac{d S(t)}{dt}&=&h N(t)-\mu S(t)-\beta S(t) I(t)+\rho R(t),\\
	& & \\
	\frac{d E(t)}{dt}&=&\beta S(t) I(t)-(\mu +\varepsilon)E(t),\\
	& & \\
	\frac{d I(t)}{dt}&=&\varepsilon E(t)-(\mu + \tau +\gamma )I(t),\\
	& & \\
	\frac{d R(t)}{dt}&=&\gamma  I(t)-(\mu + \rho)R(t).
	\end{array}\right.
\end{equation}

The description of  parameters and variables is the following

\begin{enumerate}
	\item[] \textbf{Variables}
	\begin{enumerate}
		\item[] $S(t)$: susceptible humans,
		\item[] $E(t)$: exposed humans to tuberculosis, \item[] $I(t)$: infected humans with active tuberculosis,  
		\item[] $R(t)$: recovered humans,
		\item[] $N(t)$: total population.
	\end{enumerate}
	\item[] \textbf{Parameters}
	\begin{enumerate}
		\item[] $h$: human birth rate 
		\item[] $\beta$: rate at which the susceptibles become exposed to MTB
		\item[] $\rho$: progression rate from recovered to susceptible class
		\item[] $\varepsilon$: progression rate from latent to infectious class
		\item[] $\gamma$:  progression rate from recovered to susceptible class
		\item[] $\mu$: natural human death rate
		\item[] $\tau$: human  death rate caused by the MTB tuberculosis 
	\end{enumerate}
\end{enumerate}

\subsection{Existence and Positivity of Solutions}

The system \ref{e1} is epidemiologically and mathematically well-posed in the feasible region $\Gamma$ given by
	
	\begin{equation}\label{r}
		\Gamma =\{(S, L, I, R) \in \mathbb{R}_{+}^{4}:~0 < \mu \leq h\}.
	\end{equation}
	
	\begin{prop}
		The system (\ref{e1}) always admits positive solutions for all positive
		initial conditions and the biological region $\Gamma \in \mathbb{R}_{+}^{4}$ is positively invariant and
		globally attractive for the same system .
	\end{prop}
 
\textit{Proof}:\\
\\
From the system \ref{e1}, we have
\begin{equation}\label{ee1}
\left\lbrace 
\begin{array}{ccl}
\frac{d S(t)}{dt}&\geq&-(\mu +\beta I(t))S(t),\\
& & \\
\frac{d E(t)}{dt}&\geq&-(\mu +\varepsilon)E(t),\\
& & \\
\frac{d I(t)}{dt}&\geq&-(\mu + \tau +\gamma )I(t),\\
& & \\
\frac{d R(t)}{dt}&\geq&-(\mu + \rho)R(t).
\end{array}\right.
\end{equation}
Integrating each inequality of the system \ref{ee1} yields
\begin{equation}\label{ee2}
\left\lbrace 
\begin{array}{ccl}
 S(t)&\geq&S(0) \exp \{ -(\mu t + \beta \int_{0}^{n} I(n) dn) \},\\
& & \\
 E(t)&\geq&E(0)\exp\{ -(\mu +\varepsilon)t\},\\
& & \\
I(t)&\geq&I(0) \exp\{ -(\mu + \tau +\gamma )t\},\\
& & \\
 R(t)&\geq&R(0) \exp\{-(\mu + \rho)t\}.
\end{array}\right.
\end{equation}

where $S(0), E(0), I(0), R(0)$ are all positive initial conditions. From the system \eqref{ee2}, all state variables [that means ($S(t), E(t), I(t), R(t)$) ] are all positive for all $t\geq 0$. Thus, the solutions of model \ref{e1} remain positive in $\Gamma$ for all time $t \geq 0$.
\\
\\
Add member to member the equations of the system \eqref{e1}, we notice that the total of the
human population satisfies the following relation:

\begin{equation}\label{e2}
	\dot{N}(t)= h N(t)-\mu N(t)-\tau I(t).
\end{equation}

We have $\dot{N}(t)\leq h N(t)-\mu N(t)$ according the equation \ref{e2}. Then, $\dot{N}(t)\leq 0$ if and only if $h \leq \mu$. So, $N  (t) \leq N(0) \exp[(h - \mu )t]$.  We conclude that the region $\Gamma$ is positively invariant. Moreover, if $h < \mu$, we see that the solution of the
system \ref{e1} enter in $\Gamma$ in finite time, that means $N (t)$ asymptotically approaches $h$.

Therefore, all solutions in $\mathbb{R}_{+}^{4} $ eventually enter $\Gamma$ that is the biological region $\Gamma$ is globally attractive for the same system. So the problem is then mathematically and epidemiologically well posed. Hence, every solution of the model \ref{e1} with initial conditions in $\Gamma$ remains in $\Gamma$ for all $t \geq 0$.\hfill $
\Box$

\subsection{Disease Equilibria points}

\subsubsection{Disease Free Equilibrium (DFE)}

A disease free equilibrium point is a solution of the system \eqref{e1} in holding that there is no disease in population. In this case $E = I = R = 0$. Therefore in our case the DFE, $E_0$, is expressed as
			
\begin{equation}\label{e3}
  E_0=\left(\frac{N h}{\mu},0,0,0 \right).
\end{equation}

\subsubsection{Existence of Endemic Equilibrium (EE)}

Using the definition of an equilibrium point and doing the substitutions the endemic equilibrium, $E_1$, is expressed as 
		 
\begin{equation}\label{e4}
   E_1=(S^{*}, E^{*}, I^{*}, R^{*}).
\end{equation}
		 
   where
		
\begin{equation}
		\left\lbrace 
			\begin{array}{lll}
			S^{*}&=&\frac{(\mu + \varepsilon)(\mu+\tau+\gamma)}{\beta \varepsilon},\\
			& & \\
			E^{*}&=&\frac{(\mu+\rho)(\mu+\tau+\gamma)[h N \beta \varepsilon-\mu(\mu+\varepsilon)(\mu+\tau+\gamma)]}{\beta \varepsilon[(\mu+\rho)(\mu+\varepsilon)(\mu+\tau+\gamma)-\gamma \varepsilon \rho]},\\
			& & \\
			I^{*}&=& \frac{(\mu+\rho)[h N \beta \varepsilon-\mu(\mu+\varepsilon)(\mu+\tau+\gamma)]}{\beta[(\mu+\rho)(\mu+\varepsilon)(\mu+\tau+\gamma)-\gamma \varepsilon \rho]},\\
			& & \\
			R^{*}&=&\frac{\gamma [h N \varepsilon \beta - \mu (\mu+\varepsilon)(\mu+\tau+\gamma)]}{\beta[(\mu+\rho)(\mu+\varepsilon)(\mu+\tau+\gamma)-\gamma \varepsilon \rho]}.	
			\end{array} \right.
\end{equation}

Therefore, the endemic equilibrium (EE), denoted $E_1$, given by equation \eqref{e4} exists whenever the associated reproduction threshold quantity, $R_0$, exceeds unity. In other hand, no endemic equilibrium.

\subsection{Basic Reproduction Number}

The reproduction number, $R_0$, which measures the spread of infections in a population, was computed by next generation matrix approach\cite{T14, T15}.  Mathematically, $R_0$, is a treshold for stability of a disease-free equilibrium (DFE) and is related to the peak and final size of an epidemic. In other hand, $R_0$ is defined as the average number of new cases  an infection caused by one typical infected individual, in a population consisting of susceptibles only \cite{T15}.\\
\\
Using this approach, the basic reproduction number, $R_0$, is given by $R_0=\rho(-F V^{-1})$, where 

\begin{equation}
	\begin{array}{ccc}
	F&=&\begin{pmatrix}
		0& \frac{\beta h N}{\mu}\\
		 & \\
		 0&0
	\end{pmatrix}
 
	\end{array}
\end{equation}

and

\begin{equation}
	\begin{array}{ccc}
	V^{-1}&=&\begin{pmatrix}
	\frac{-1}{\mu + \varepsilon}& 0\\
	& \\
	\frac{-\varepsilon}{(\mu + \varepsilon) (\gamma + \mu + \tau)}&\frac{-1}{\gamma + \mu + \tau}
	\end{pmatrix}
	\end{array}.
\end{equation}

Which means that $R_0$ is given by dominant eigenvalues of $-F V^{-1}$, where $F$ is the transmission part, describing the production of new infections, and the $V^{-1}$ is the inverted matrix of $V$ where $V$ is the transition part, describing changes in state. Thus,

\begin{equation}\label{re}
	R_0 = \frac{\beta N h \varepsilon}{\mu (\mu + \varepsilon) (\gamma + \mu + \tau)}.
\end{equation}

\subsection{Stability Analysis of disease equilibria}


\subsubsection{Stability of DFE} 

\begin{theo}
\label{th1}
The disease free equilibrium $E_0$ of the model \ref{e1}, given by equation \ref{e3}, is locally asymptotically stable (LAS) if $R_0 \leq 1$ and unstable if  $R_0 > 1 $.
\end{theo}

\textit{Proof}: This theorem will be proved based on the notions of the Jacobian matrix. Let then be a point $M=(S, E, I, R)$ and $J(M)$ its Jacobian. We have:

\begin{equation}
	J(M)= \begin{pmatrix}
	-(\mu + \beta I)& 0 & - \beta S & \rho\\
	& & & \\
	\beta I&-(\mu + \varepsilon)& \beta S & 0\\
	& & & \\
	0& \varepsilon & -(\mu +\tau + \gamma)& 0\\
	& & & \\
	0 & 0 & \gamma & -(\mu + \rho)
	\end{pmatrix}.
\end{equation}

The Jacobian of DFE, $E_0$, expressed in equation \ref{e3}. Finding the eigenvalues, $z$, of $J(E_0)$  amounts to solving the equation 

\begin{equation}\label{e5}
	P(z) = 0.
\end{equation}

 where $P(z)$ is the characteristic polynomial of $J(E_0)$, that is\\ $P(z)=|J(E_0) - z \mathbb{I}_4|$. 
 
 The equation \ref{e5} becomes:

\begin{equation}
	(\mu + z)=0,
\end{equation}

or

\begin{equation}
    (\mu + \rho+z)=0
    \end{equation}
    or
    \begin{equation}
z^2 + (2 \mu + \varepsilon + \tau + \gamma)z + (\mu + \varepsilon) - \frac{h \beta \varepsilon N}{\mu} = 0
\end{equation}

Then, we have:\\
\begin{equation}\label{e6}
	 z_1 = - \mu,
\end{equation}
 or
 \begin{equation}\label{e7}
 	z_2 = - (\mu + \rho),
 \end{equation}
 or
 \begin{equation}\label{e8}
 	z^2 + (2 \mu + \varepsilon + \tau + \gamma)z + (\mu + \varepsilon) (\mu + \tau + \gamma) - \frac{h \beta \varepsilon N}{\mu} =0,
 \end{equation}
 
with Descartes' rule of signs\cite{T17}, if $R_0 \leq 1$, all the coefficients of the polyom characterizing the left side of the equation \ref{e8}
are strictly positive, then it does not have a positive root. From \ref{e6} \ref{e7} \ref{e8}, if $R_0 \leq 1$, $J(E_0 )$
has all its eigenvalues with strictly negative real part, hence $E_0$ is locally
asymptotically stable (LAS) by the Poincarré-Lyapunov theorem\cite{T17, T18}. 

Otherwise,
with the same rule, if $R_0 > 1$, there is a variation of coefficients of $P (z)$, then $P (z)$ has at least one positive root. So, $J(E_0 )$ has at least one eigenvalue with a strictly positive real part, hence then $E_0$ is unstable by the Poincarré-Lyapunov theorem. \hfill $\Box$\\

The epidemiological implication of theorem \ref{th1} is that Tuberculosis spread can be effectively controlled in the community when $R_0 \leq 1$ that means if the initial sizes of the populations of the model are in the basin of attraction of the disease free equilibrium $E_0$. To ensure that elimination of TB is independent of the initial sizes of the populations, it is necessary to show that the DFE is globally asymptotically stable\cite{T19}. It is shown below that tuberculosis will be eliminated from the community if the epidemiological threshold can be reduced to a value below unity.

\begin{theo}\label{th2}
The disease free equilibrium $E_0$ of the model \ref{e1}, given by equation \ref{e3}, is globally asymptotically stable (GAS) in $\Gamma$ whenever $R_0 \leq 1$ and unstable if  $R_0 > 1 $.
\end{theo}

 \textit{Proof}: Consider the following Lyapunov candidate function for $E_0$:
	
	\begin{equation}
		L=\varepsilon E + (\mu + \varepsilon) I + A R~~\text{with}~A= \frac{\varepsilon \beta N (h- \mu)}{\gamma \mu}.
	\end{equation}
	The first derivative of $L$ along the solutions of model \ref{e1} is
	
	\begin{equation}
		\begin{array}{cll}
		\dot{L}&=&\varepsilon [\beta S I - (\mu + \varepsilon)E] +(\mu + \varepsilon) [\varepsilon E - (\mu +\tau + \gamma) I] + A [\gamma I - (\mu +\rho) R]\\
		 & & \\
		 & \leq & [\varepsilon \beta S - (\mu + \varepsilon) (\mu +\tau + \gamma) + A \gamma ]I\\
		 & & \\
		 & \leq &  (\mu + \varepsilon) (\mu +\tau + \gamma) (R_0 - 1) I.
		\end{array}
	\end{equation}
	Since all the model parameters are non negative, it follows that $\dot{L} \leq 0$ for $R_0 \leq 1$. For $R_0 = 1$, $\dot{L}=0$ if and only if $E=I=R=0$. As $t \longrightarrow \infty$, substituting these values in model \ref{e1}, we have that $S \longrightarrow \frac{ N h}{\mu},~ E \longrightarrow 0,~ I \longrightarrow 0,~\text{and}~R \longrightarrow 0 $. Hence, $L$ is a Lyapunov candidate function on $\Gamma$, and the largest
	compact invariant set in $\{( S ,  E , I , R ) \in \Gamma : \dot{L} = 0 \}$ is the singleton $\{ E_0 \}$. Therefore, by LaSalle's Invariance Principle \cite{T19,T20}, every solution of model \ref{e1}, with initial conditions in $\Gamma$, approaches $E_0$ as $t \longrightarrow \infty$, whenever $R_0 \leq 1$ that is the disease free equilibrium $E_0$ of the model \ref{e1}, given by equation \ref{e3}, is globally asymptotically stable (GAS) in $\Gamma$ whenever $R_0 \leq 1$ and unstable if  $R_0 > 1 $. \hfill $\Box$

\subsubsection{Stability of EE}

\begin{theo}\label{th3}
	The endemic equilibrium $E_1$ of the model \ref{e1}, given by equation \ref{e4}, is locally asymptotically stable (LAS)  if  $R_0 > 1 $.
\end{theo}
\textit{Proof}: From $N=S+E+I+R$, we deduce that $R =N-S-E-I$. By reducing dimension of system \ref{e1} while keeping the variables $S, E, I$, we have:

\begin{equation}\label{e9}
\left\lbrace 
\begin{array}{ccc}
\frac{d S(t)}{dt}&=&h N(t)-\mu S(t)-\beta S(t) I(t)+\rho (N-S-E-I),\\
& & \\
\frac{d E(t)}{dt}&=&\beta S(t) I(t)-(\mu +\varepsilon)E(t),\\
& & \\
\frac{d I(t)}{dt}&=&\varepsilon E(t)-(\mu + \tau +\gamma )I(t).
\end{array}\right. .
\end{equation}
Let $W=(S, E, I)$ be a point and $J(W)$ its Jacobian. We have:

\begin{equation}
J(W)= \begin{pmatrix}
-(\mu +\rho + \beta I)& -\rho& -(\rho+\beta S) \\
& & & \\
\beta I&-(\mu + \varepsilon)& \beta S \\
& & & \\
0& \varepsilon & -(\mu +\tau + \gamma)\\
\end{pmatrix}.
\end{equation}
In this form, the reduction of endemic equilibrium point EE, $E_1$, becomes $E_1 = (S^{*}, E^{*}, I^{*})$.
Let us look for the eigenvalues of $ J(E_1)$, this amounts to find the roots of the characteristic polynomial $P(z)$ that means resolve $P(z)=0$.
By factoring determinant above, we have

\begin{equation}
	P(z)=A(z) B(z) C(z)
\end{equation}
with
\begin{equation}\label{e10}
	A(z)=-[(\mu +\rho + \beta I^{*}) +z],
\end{equation}

\begin{equation}\label{e11}
	B(z)=z^2 +(2 \mu + \varepsilon + \rho +\beta I^{*})z+(\mu+\rho + \beta I^{*})(\mu + \varepsilon)+ \rho \beta I^{*},
\end{equation}
and
\begin{equation}\label{e12}
	\begin{array}{lll}
	C(z)&=& z^4 +(3 \mu + 2 \beta I^{*} + 2 \rho + \varepsilon +\tau +\gamma) z^3\\
	& & \\
	& & +[(2 \mu +\rho + \beta I^{*} +\tau +\gamma) (2 \mu +\varepsilon + \rho + \beta I^{*}) \\
	& &\\
	& &+ (2 \mu +\rho + \beta I^{*}) (\mu + \varepsilon)+ (\mu + \rho +\beta I^{*}) (\mu + \tau + \gamma) + \rho \beta I^{*}] z^2\\
	& & \\
	& &+[(\mu + \rho +\beta I^{*}) (\mu + \tau + \gamma) (2 \mu + \varepsilon + \rho + \beta I^{*}) +\varepsilon \beta S^{*}]z\\
	& & \\
	& &+[(\mu + \rho + \beta I^{*}) (\mu + \varepsilon)+\rho +\beta I^{*}] [2 \mu + \rho +\beta I^{*} \tau + \gamma\\
	& &\\
	& &+(\mu +\rho + \beta I^{*}) (\mu + \rho +\beta I^{*}) (\mu +\tau +\gamma)] \\ & &+ \varepsilon \beta S^{*} (\mu + \rho +\beta I^{*}) +\rho \beta I^{*}(\rho + \beta S^{*})
	\end{array}.
\end{equation}

Resolve $P(z)=0 \Longrightarrow A(z)=0,~~\text{or}~~B(z)=0,~~\text{or}~~C(z)=0$. Thus, with Descartes' rule of signs\cite{T17}, if $R_0 > 1$, all the coefficients of the polyom characterizing the right side of the equations \ref{e11}, \ref{e12}
are strictly positive, then they do not have a positive root. From \ref{e10} \ref{e11} \ref{e12}, if $R_0 > 1$, $J(E_1 )$
has all its eigenvalues with strictly negative real part, hence $E_1$ is locally asymptotically stable (LAS) by the Poincarré-Lyapunov theorem\cite{T17, T18}. Therefore, this is the end of the theorem \ref{th3}. \hfill $\Box$

This implies that the tuberculosis persist and invade the population if $R_0 > 1$.

\section{Numerical simulation}

In this section, we are going to demonstrate the behavior of the proposed TB infection model through numerical simulations while comparing the model with actual tuberculosis data in Burundi. Data from Burundi were collected on quarterly (resp. annually) from the Ministry of Public Health between 2011 and 2019 (resp. between 1997 and 2020). As shown in Table \ref{my_table}, the parameter values for simulation are either from literature or reasonably chosen estimates. We used  Python to run simulations and the Runge-Kutta method was used to approximate solutions of the SEIR-system. 

\begin{table}[H]
    \caption{Parameters values}
    \centering
    \begin{tabular}{c|p{9cm}}

         \hline
	\textbf{Parameter}&\textbf{Parameter Value and references}\\
	\hline
	  h & Human birth rate is 38.377 per year per 1000 inhabitants \cite{T18} \\
	  \hline
	$\beta$ & 0.9 Probability of being infected for a susceptible meted by an infectious  \cite{T23} \\
	\hline
	$\rho$ & 0.897 Rate of progression from recovered to susceptible state \textit{Estimated}\\
	\hline
	$\gamma$ & 0.525 Rate of progression from infectious to recovered state \textit{Estimated}\\
	\hline
	$\mu$ &  Human (natural) mortality rate is 7.766 per year per 1000 inhabitants \cite{T18}\\
	\hline
	$\varepsilon$ &0.25 Rate of progression from exposed to infectious state \cite{T24} \\
	\hline
	$\tau$ & 21.0 (per 100,000 people) in 2019 human tuberculosis-induced death rate \cite{T25} \\
	\hline
 
    \end{tabular}
    \label{my_table}
\end{table}

\subsection{Fitting the model to tuberculosis data in Burundi}

In this paragraph, we use systems \ref{e1} and  parameters listed in the table \eqref{my_table} to fit the tuberculosis model to the annual  data of population which are diagnostics of tuberculosis in Burundi from $1985$ to $2020$. 
So, $S_0 =  8,895,836,~E_0 =   375,~I_0 =   1789$ and $R_0 =   60000 $ are assumed initial conditions of states 
	
\begin{figure}[H]
			\begin{centering}
					\begin{subfigure}{.450\textwidth}
						\includegraphics[width=\linewidth]{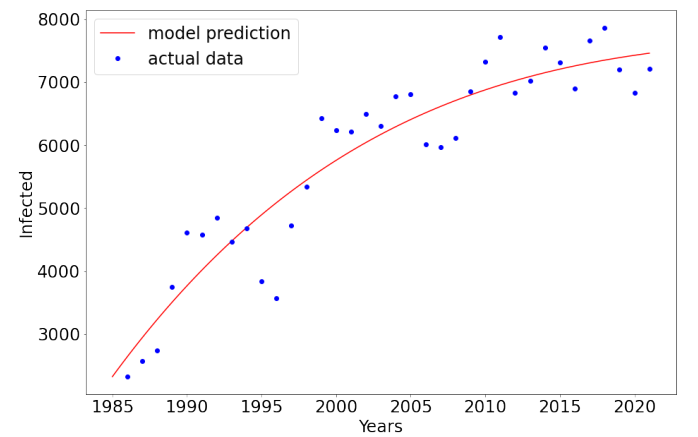}
					\end{subfigure}%
				\begin{subfigure}{.450\textwidth}
						\includegraphics[width=\linewidth]{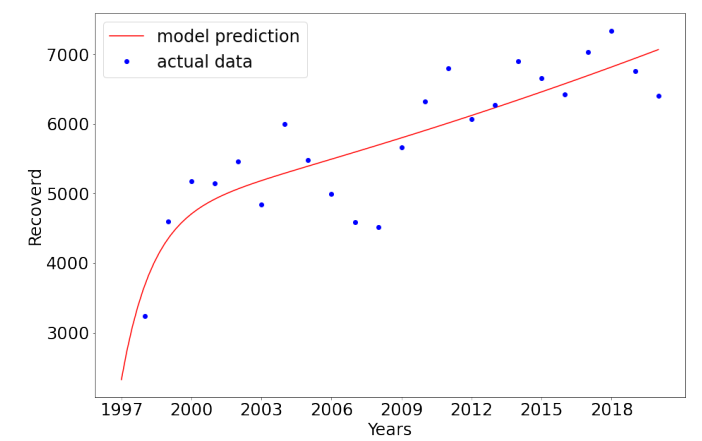}
					\end{subfigure}%
\end{centering}
\caption{Fitting the model to tuberculosis data in Burundi}
\label{recovered}
\end{figure}

The yearly data in Figure \eqref{recovered} follows the trend more closely. In this case, we have obtained a good fit. Additionally, the yearly increase in TB cases could be attributed to poor interventions implementation.

\subsection{Simulation of TB transmission dynamics}

\begin{figure}[!h]
		\begin{centering}
			\begin{multicols}{2}
				\begin{subfigure}{.54\textwidth}
		\includegraphics[width=\linewidth]{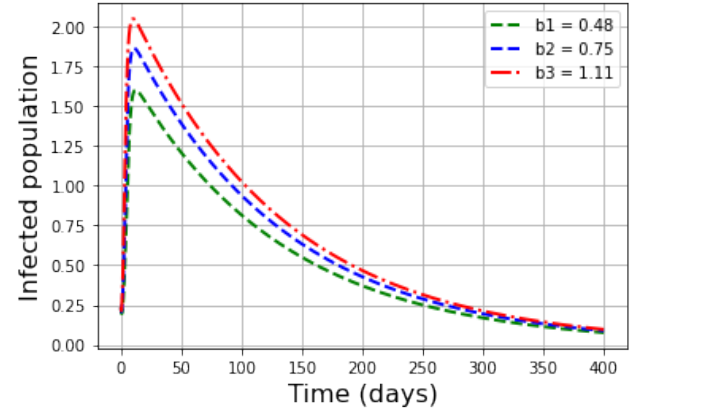}
				\end{subfigure}%
				\begin{subfigure}{.50\textwidth}
		\includegraphics[width=\linewidth]{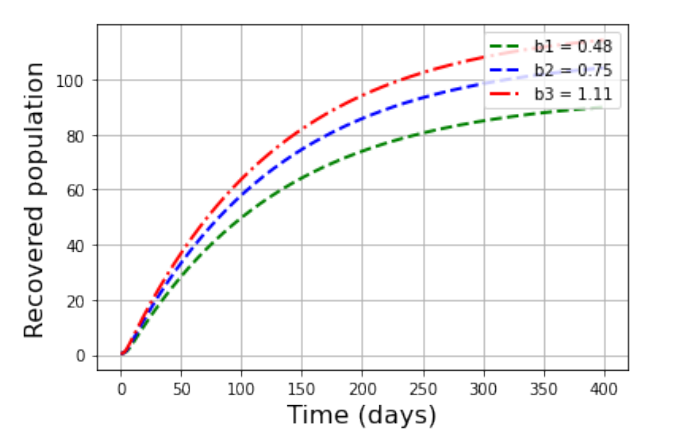}
				\end{subfigure}%
		\end{multicols}
	\end{centering}
\caption{Human population dynamics in relation to infectious, and recovered humans when  $R_0>1$}
\label{ro}
\end{figure}

Figure \eqref{ro} demonstrates that if $R_{0}$ is more than 1, each infectious person spreads the disease to more than one new person throughout the time of contagion, which results in the spread of the disease throughout the population and the emergence of an epidemic

\begin{figure}[!h]
		\begin{centering}
			\begin{multicols}{2}
				\begin{subfigure}{.50\textwidth}
		\includegraphics[width=\linewidth]{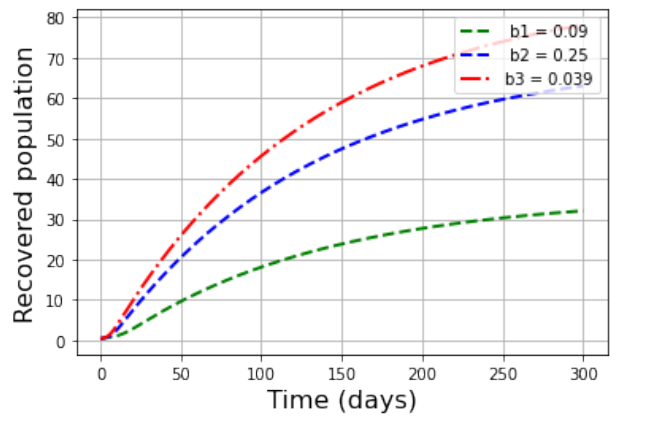}
				\end{subfigure}%
				\begin{subfigure}{.53\textwidth}
		\includegraphics[width=\linewidth]{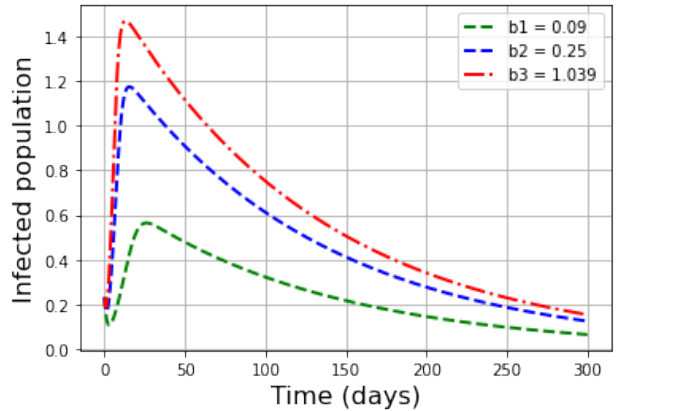}
				\end{subfigure}%
		\end{multicols}
	\end{centering}
\caption{Human population dynamics in relation to infectious, and recovered humans when $R_0<1$}
\label{r1}
\end{figure}

and Figure \eqref{r1} shows that if $R_{0}$ less than 1 suggests that each person infects, on average, less than one new person, resulting in disease extinction.

\newpage
\section{Sensitivity Analysis}

In order to determine the relative importance of model parameters on the disease infection, a sensitivity of parameters of the model system \ref{e1} is carried out. Sensitivity indices allow us to measure the relative change in a state variable when a parameter changes \cite{T21}. The numerical calculation of the sensitivity indices also allows the determination of parameters which have a reasonable impact on the basic reproduction number $R_0$ and which of the parameters is most sensitive which can help in the eradication of the disease in the population \cite{T22}.

The normalized forward sensitivity index of a variable to a parameter is the ratio of the relative change in the variable to the relative change in the parameter. When the variable is a differentiable function of the parameter, the sensitivity index may be alternatively defined using partial derivatives \cite{TA, T21, T22}.

\begin{de}
The normalized forward sensitivity index of a variable, $u(p)$, that depends differentiably on a parameter, $p$, is defined as:
	
\begin{equation}
    K_{p}^{u} = \frac{\partial u}{\partial p} \times \frac{p}{u},~~~\text{with}~~u \neq 0
\end{equation}
\end{de}

Therefore, we derive an analytical expression for the sensitivity of $R_0$ with $R_0$ given by the expression \ref{re}. Thus,

\begin{equation}\label{re1}
K_{p_i}^{R_0} = \frac{\partial R_0}{\partial p_i} \times \frac{p_i}{R_0},~~~\text{with}~~p_i : i =1, \cdots, n
\end{equation}
$p_i$ denotes each parameter involved in $R_0$.

According to Table \ref{my_table}, and the equation \ref{re1}, we plot sensitivity index of each parameter with respect to the $R_0$ given by equation \ref{re}. Sensitivity indices on $R_0$  are showed on the the following bar chart

\begin{figure}[!h]
	\centering
	\includegraphics[scale=0.55]{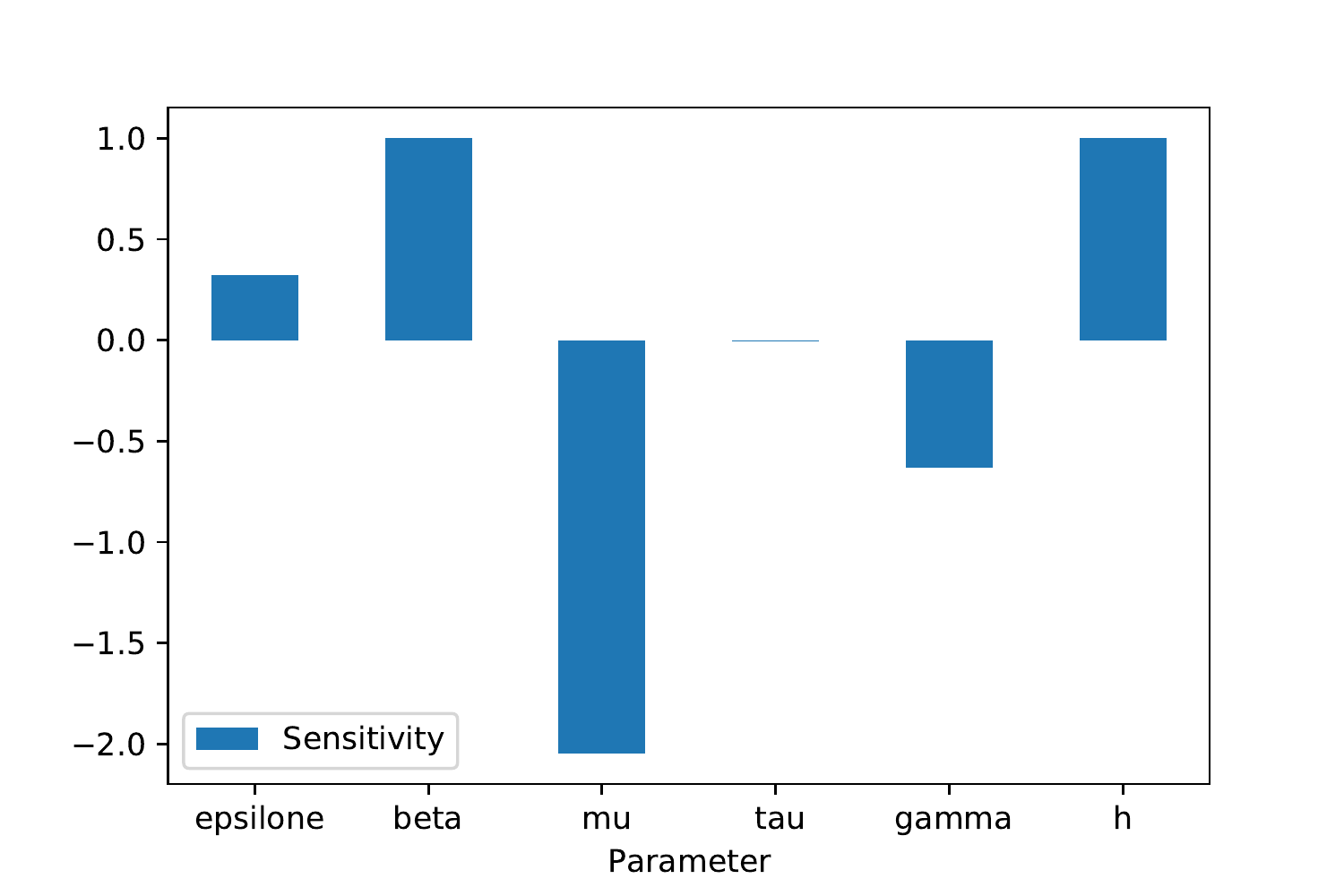}
	\caption{Bar chart for sensitivity}
 \label{f2}
\end{figure}

All parameters of the model are assumed to be non-negative, let the total population be 100,000. Hence most sensitive parameter is being the highest positive index. On Figure \ref{f2} we can list $\beta, ~h~\text{and}~\varepsilon$. This implies that increasing these parameters causes $R_0$ to increase by $100\%$. Thus, since $R_0$ continues to be higher, the epidemic of disease infection tends to occur. On the other hand, the parameters whose sensitivity is negative ($\gamma,~\tau~\text{and}~\mu$) have a highest negative impact on $R_0$. For the following, a descriptive analysis as well as the linear adjustment of our model will be carried out.

\section{Conclusion}

In this paper, a deterministic mathematical model for the transmission dynamics of TB in Burundi has been carried out to study the
stability of both disease-free and endemic equilibrium point. Using matrix
generation approach, the basic reproduction number $R_0$ was computed.
Therefore, the disease-free equilibrium of the model obtained is both locally and globally stable for $R_0 \leq 1$. It is also shown that the endemic equilibrium solution of the model is globally asymptotically stable if $R_{0} > 1$. Sensitivity Analysis has showed that $\beta, ~h~\text{and}~\varepsilon$ are being the highest positive index.

Using data from Ministry of Public Health and the Fight against
AIDS, parameters of the model have been identified and the model is shown to describe TB dynamic in Burundi from 1997 to 2019 (annual data) and from 2011 to 2019 (quarterly data).      

The results of this research support the idea that in sub-Saharan Africa people in general and in Burundi in particularly should be strongly encouraged to seek a diagnosis of TB, and the success rate the treatment must be correlated with the population of infectious diagnosed. Then,
the population of diagnosed infectious diseases and the number of TB-related deaths will decrease. 

\bibliographystyle{unsrt}
\bibliography{bibliography.bib}

\end{document}